\documentclass[submission, Proceedings]{SciPost}

\pdfoutput=1

\hypersetup{
    colorlinks,
    linkcolor={red!50!black},
    citecolor={blue!50!black},
    urlcolor={blue!80!black}
}

\usepackage[bitstream-charter]{mathdesign}
\usepackage{caption}
\usepackage{subcaption}
\usepackage{wrapfig}

\DeclareSymbolFont{usualmathcal}{OMS}{cmsy}{m}{n}
\DeclareSymbolFontAlphabet{\mathcal}{usualmathcal}

\def\Xmax{$X_{\rm{max}}$}

\begin{document}

\begin{center}{\Large \textbf{
Probing hadronic interactions with measurements from the Pierre Auger Observatory
\\
}}\end{center}

% TODO: write the author list here. Use initials + surname format.
% Separate subsequent authors by a comma, omit comma at the end of the list.
% Mark the corresponding author with a superscript *.
\begin{center}
B. Andrada\textsuperscript{1}
on behalf of the Pierre Auger Collaboration\textsuperscript{2$\star$}
\end{center}

\begin{center}
{\bf 1} Instituto de Tecnologías en Detección y Astropartículas (CNEA, CONICET, UNSAM), Buenos Aires, Argentina
\\
{\bf 2} Observatorio Pierre Auger, Av. San Martín Norte 304, 5613 Malargüe, Argentina\\
(Full author list: \href{https://www.auger.org/archive/authors_2021_07.html}{https://www.auger.org/archive/authors\_2021\_07.html})
\\
* spokespersons@auger.org
\end{center}

\begin{center}
\today
\end{center}

% For convenience during refereeing (optional),
% you can turn on line numbers by uncommenting the next line:
%\linenumbers

\definecolor{palegray}{gray}{0.95}
\begin{center}
\colorbox{palegray}{
  \begin{tabular}{rr}
  \begin{minipage}{0.1\textwidth}
    \includegraphics[width=30mm]{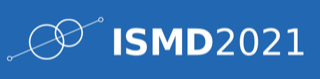}
  \end{minipage}
  &
  \begin{minipage}{0.75\textwidth}
    \begin{center}
    {\it 50th International Symposium on Multiparticle Dynamics}\\ {\it (ISMD2021)}\\
    {\it 12-16 July 2021} \\
    \doi{10.21468/SciPostPhysProc.?}\\
    \end{center}
  \end{minipage}
\end{tabular}
}
\end{center}

\section*{Abstract}
{\bf
% TODO: write your abstract here.
%The abstract is in boldface, and should fit in 8 lines.
%It should be written in a clear and accessible style, emphasizing the context, the problem(s) studied, the methods used, the results obtained, the conclusions reached, and the outlook. You can add a table contents, recommended if your paper is more than 6 pages long.
The Pierre Auger Observatory is the largest facility in the world to study ultra-high-energy cosmic rays. It has a hybrid detection technique that combines the observation of the longitudinal development of extensive air showers and the measurement of their particles at the ground. This capability has opened the possibility to probe hadronic interactions taking place at energies well beyond those accessible by human-made accelerators. In this report, we present a selection of the latest results on hadronic interactions with measurements from the Pierre Auger Observatory. These data span over three decades in energy, showing the tension between data from the muon component of air showers and predictions based on the most updated hadronic interaction models.
}

%\vspace{10pt}
%\noindent\rule{\textwidth}{1pt}
%\tableofcontents\thispagestyle{fancy}
%\noindent\rule{\textwidth}{1pt}
%\vspace{10pt}

\section{Introduction}
\label{sec:intro}
% TODO: write your article here.
% stage is yours. Write your article here.
%The bulk of the paper should be clearly divided into sections with short descriptive titles, including an introduction and a conclusion.
When ultra-high-energy cosmic rays reach the upper atmosphere, they interact with air nuclei at a center of mass energy up to $\sqrt{s} \sim 400$ TeV\@. Meanwhile, human-made accelerators are able to produce p-p collisions at $\sqrt{s} = 13$ TeV\@. Thus, ultra-high-energy cosmic rays provide access to hadronic interactions that are not accessible with accelerators. However, they cannot be measured directly with good statistics because of their extremely low flux. All their properties have to be inferred from the extensive air showers of secondary particles they produce when colliding with atmospheric nuclei. These showers have two components. The electromagnetic part is formed predominantly by the decay of neutral pions and grows through pair production and bremsstrahlung. The hadronic part, on the other hand, is formed mainly by chain reactions of secondary mesons and baryons which interact until their energies are low enough such that the probability for decay becomes greater than that of further interactions. Muons are one of the products of these decays. 

The Pierre Auger Observatory \cite{NIM_2015}, located near the city of Malargüe, in Argentina, is the largest facility in the world designed to observe ultra-high-energy cosmic rays and it combines two independent measurement techniques providing a hybrid detection mechanism. The Surface Detector (SD) studies the lateral distribution of air shower particles that reach ground level with more than 1600 water-Cherenkov detectors (WCD) that have a duty cycle close to 100\%. WCDs are distributed in three hexagonal grids with different spacing between detectors and covering different areas. The SD-1500 covers an area of about $3000$ km$^2$ in a hexagonal grid of $1500$ m spacing, while the SD-750 and SD-433 cover areas of $27$ km$^2$ and $1.9$ km$^2$ with hexagonal grids of $750$ m and $433$ m spacing respectively. Each of these arrays aims at covering different primary energies: the larger the spacing and area covered, the higher the observed energy. The Fluorescence Detector (FD) views the atmosphere above the SD with $27$ telescopes distributed in four sites surrounding the SD. It measures the nitrogen fluorescence light produced by the passage of relativistic charged particles through the atmosphere and operates only on clear and moonless nights, having a $\sim 15\%$ duty cycle. A major upgrade is taking place in the Auger Observatory \cite{Prime_2016}, AugerPrime, that will enable the disentanglement of the electromagnetic and muonic components of the shower, improving the sensitivity to hadronic interactions and mass composition. This upgrade includes the installation of an Underground Muon Detector (UMD) to allow for the direct detection of the muon content of extensive air showers. This detector consists of 30 m$^2$ scintillators located next to each WCD of the SD-750 and buried 2.3 m underground to shield all particles of the shower except for muons with energies greater than 1 GeV\@.  

One of the main purposes of the Pierre Auger Observatory is to understand the mass composition of ultra-high-energy cosmic rays and its astrophysical implications \cite{MassComp_2020}. An important observable involved in these studies is the atmospheric depth of maximum development of the shower, \Xmax, which is sensitive to mass composition because showers from heavier primaries develop higher in the atmosphere (lower \Xmax) and their profiles fluctuate less than those of showers produced by lighter primaries. \Xmax\, and its fluctuations can be directly measured with the FD. Further information relevant to mass composition studies can be obtained by looking at the muon content of air showers, due to heavier primaries being expected to produce larger number of muons than lighter primaries. This is done with Auger data using different observables and techniques, coming both from SD and UMD measurements.   

The measurements of extensive air shower properties, however, need to be interpreted using the most updated hadronic interaction models which are tuned to the latest LHC data. When doing so, a deficit in the number of muons predicted by the models has been observed by a number of different experiments, using diverse detection techniques and covering a wide energy range (above $10^{16}$ eV\@) \cite{whisp_2020}. The Auger Observatory, being able to study different hadronically-sensitive observables, has contributed to determine and characterize the muon deficit in simulations with several results, a selection of which will be given here.

\section{Top-Down analysis}
\label{sec:topdown}

A novel \emph{top-down} analysis was performed on a set of high-quality hybrid events, with energies between $10^{18.8}$ eV\@ and $10^{19.2}$ eV\@. This study is based on the fact that when an attempt is made to reproduce measured events with simulations, the longitudinal profiles, as seen by the FD, can be well reproduced but the lateral distribution of particles at the ground, as measured with the SD, produce systematically smaller signals in the simulations than the ones found in the data. For each event, batches of simulations were generated using the measured energy and geometry as inputs and considering different primaries. In those cases where a matching longitudinal profile was found, the discrepancies in the SD signals were scrutinized. 

\begin{center}
  \begin{figure}
    \begin{subfigure}{0.55\textwidth}
      \includegraphics[width=0.9\textwidth]{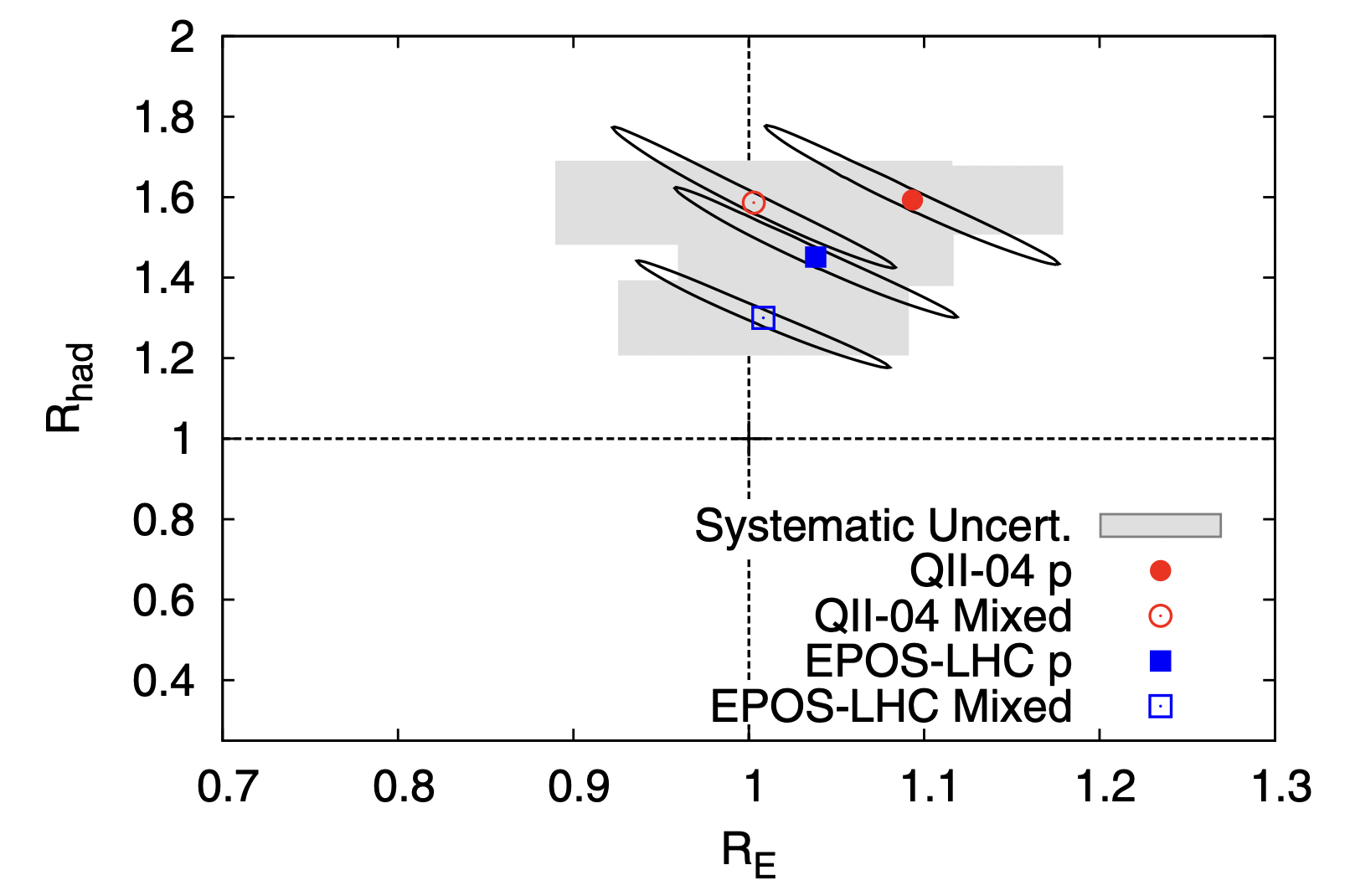}
    \end{subfigure}
    \hfill
    \begin{subfigure}{0.45\textwidth}
      \includegraphics[width=0.9\textwidth]{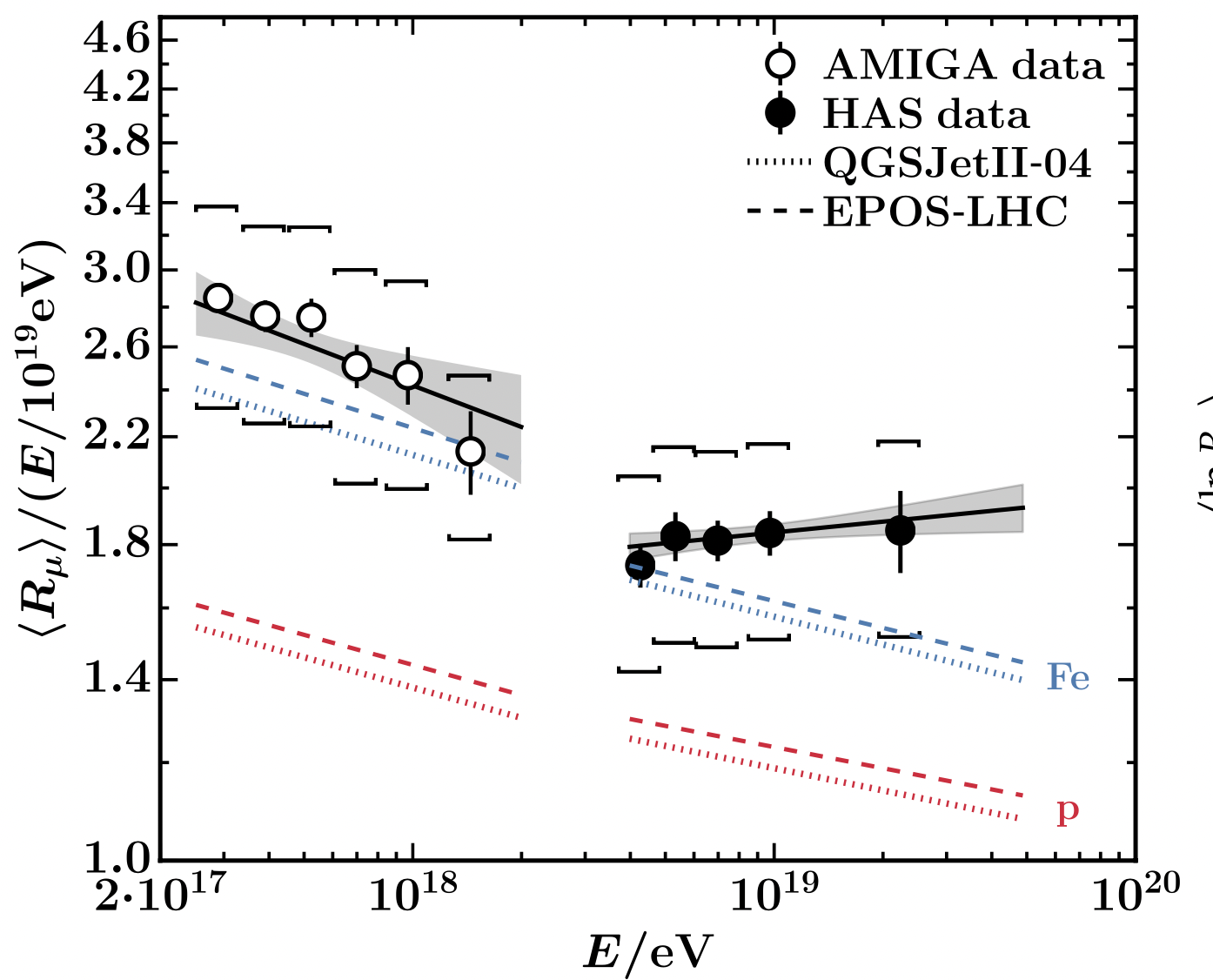}
    \end{subfigure}
    \caption{\emph{Left:} Rescaling factors for the simulated hadronic signal $R_{\rm{had}}$ and the primary energy $R_E$ for two hadronic interaction models considering alternatively a pure proton composition and a mixed proton-iron composition. The grey rectangles indicate the systematic uncertainties. \cite{TopDown_2016}.  \emph{Right:} Energy evolution of the average muon content as measured with the UMD (white circles) and with the SD (black circles). Predictions for proton and iron primaries for two hadronic interaction models are also shown. \cite{Cazon_2019}}
    \label{fig:topdown_has_umd}
  \end{figure}
\end{center}

The measured signals in the WCDs derive from both muons and electromagnetic particles and these contributions cannot be easily decoupled. However, it is straightforward to obtain both the electromagnetic and hadronic parts of the signal when using simulations. Two scaling factors were defined to allow for shifts in each component of the simulated signals: $R_E$ rescales the primary energy affecting both components of the signal and $R_{\rm{had}}$ rescales only the contribution to the ground signal of inherently hadronic origin, which consists mostly of muons. Given an appropriate set of events, these two factors can be separately determined because of the different behaviour of the components with zenith angle, namely that the electromagnetic part of the shower is more strongly attenuated in the atmosphere than the muonic and the slant depth increases with zenith angle. A maximum likelihood fit was performed to find the parameters that minimize the difference between the rescaled simulated signals and the measured ones. Results are shown in Figure \ref{fig:topdown_has_umd} (left) for two hadronic interaction models and two composition scenarios: pure proton and a mixed proton-iron composition. For the mixed composition scenario, no energy rescaling of the primary was necessary, but the hadronic signal in measurements is between $30\%$ and $60\%$ larger than the one predicted by hadronic interaction models. The details of this analysis can be found in \cite{TopDown_2016}.
 
\section{Highly inclined events}
\label{sec:has}

When the incoming cosmic ray has a large zenith angle, $\theta > 60^{\circ}$, the electromagnetic component of the corresponding extensive air shower is strongly attenuated in the atmosphere. By the time the shower front reaches ground level, the particles that produce a signal in the WCDs are mostly muons or electrons arising from the decay of muons. Thus, a data set of hybrid, high energy, highly inclined events allows for the simultaneous and independent measurement of the two shower components: the electromagnetic part of the shower is seen by the FD while the SD provides information on the muon content. 

In Figure \ref{fig:topdown_has_umd} (right) the average muon content $R_{\mu}$ (black circles) is shown as a function of the primary energy reconstructed by the FD. Predictions from hadronic interaction models for proton and iron primaries are also shown. A higher number of muons is found in data than in simulations. Even though the data is compatible within the uncertainties with the predictions for iron primaries, when independent \Xmax\, measurements (directly related with the primary composition) are contemplated in the analysis, as can be seen in Figure \ref{fig:comp} (left) for the events in the energy bin corresponding to $10^{19}$ eV\@, the measured number of muons clearly exceeds all model predictions. A thorough description of this analysis can be found in \cite{HAS_2015} and an update of it is available in \cite{Fluctuations_2021}.

\section{Direct muon measurement}
\label{sec:umd}

\begin{center}
  \begin{figure}
    \begin{subfigure}{0.48\textwidth}
      \includegraphics[width=0.9\textwidth]{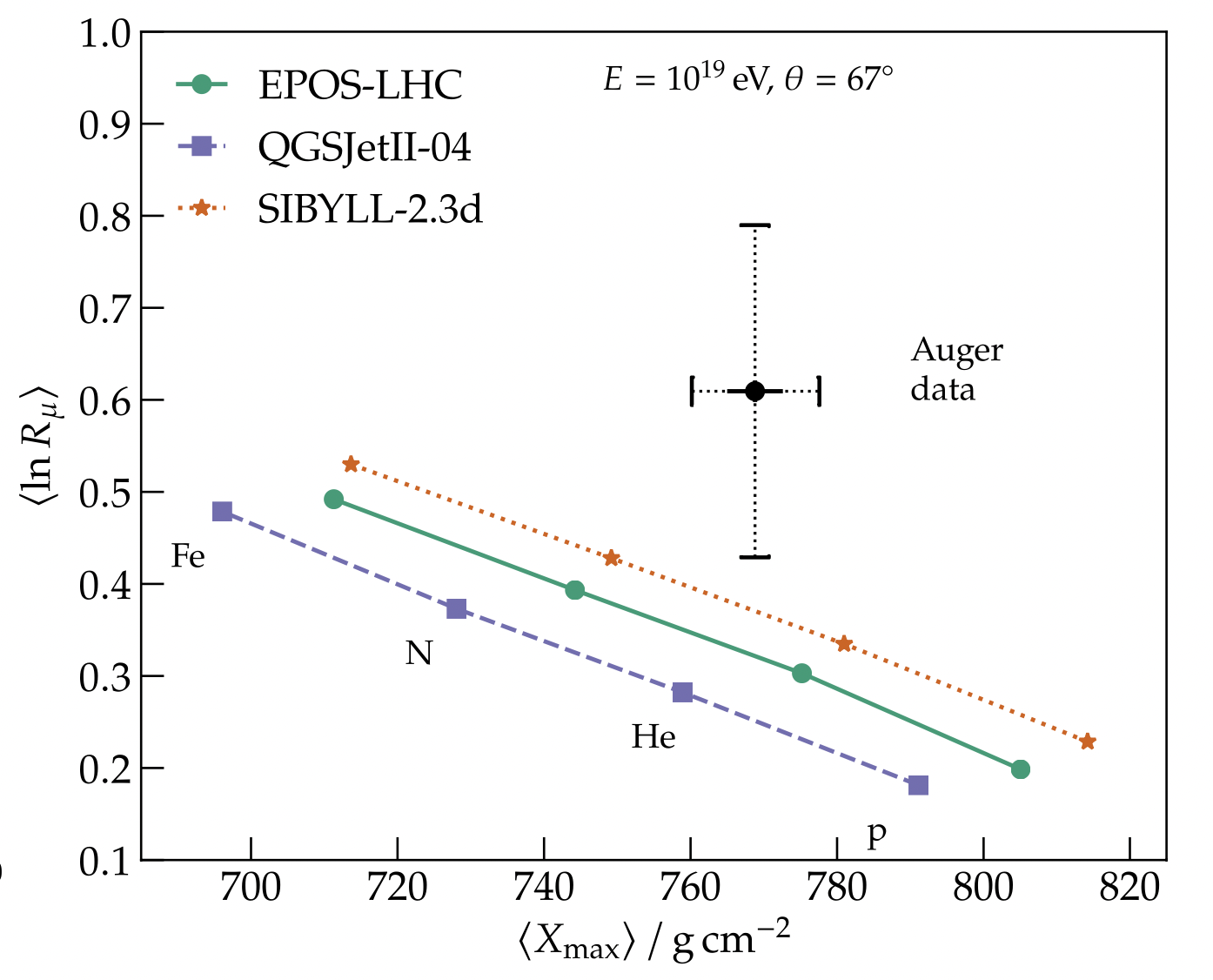}
    \end{subfigure}
    \hfill
    \begin{subfigure}{0.52\textwidth}
      \includegraphics[width=\textwidth]{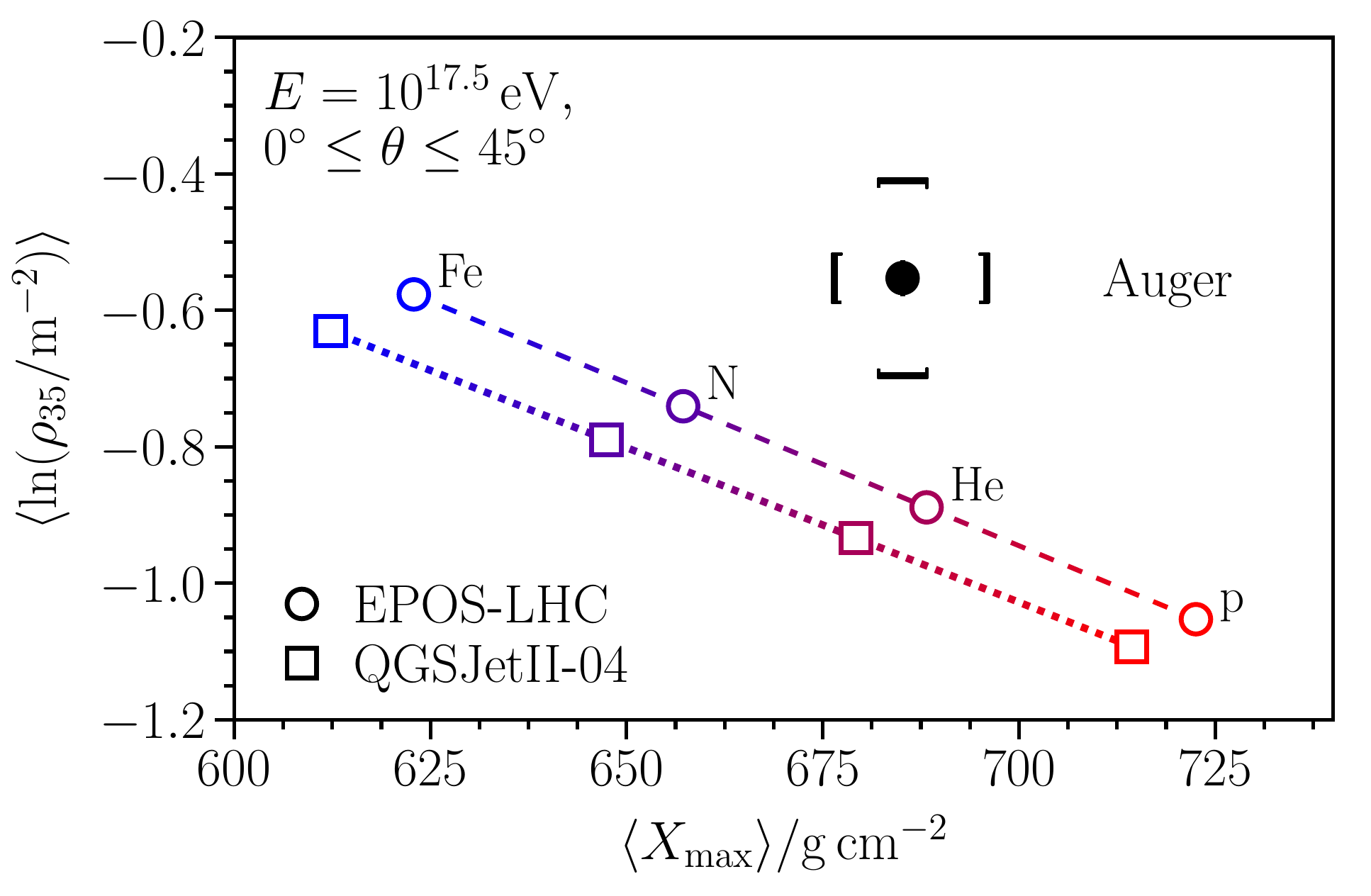}
    \end{subfigure}
    \caption{\emph{Left:} Auger SD measurement and model predictions of the average muon number and the depth of shower maximum for the events in the energy bin corresponding to $10^{19}$ eV\@ from Figure \ref{fig:topdown_has_umd} (right). \cite{Fluctuations_2021} \emph{Right:} Auger UMD measurement and model predictions of the average muon number and the depth of shower maximum for the events in the energy bin corresponding to $10^{17.5}$ eV\@ from Figure \ref{fig:topdown_has_umd} (right). \cite{UMD_2020}}
    \label{fig:comp}
  \end{figure}
\end{center}

The direct measurement of the muon content with the engineering array of the UMD provides insight into the muon deficit issue on a lower energy region, between $2\times10^{17}$ eV\@ and $2\times10^{18}$ eV\@, for the first time with Auger data. The muon content obtained with UMD data is shown in Figure \ref{fig:topdown_has_umd} (right) as open circles in the lower energy region of the plot, together with predictions from two hadronic interaction models for proton and iron primaries. Once again, predictions from the models fall short with respect to Auger data points, even though they are compatible within uncertainties with heavy primary predictions. However, when independently obtained \Xmax\, information is considered, as was done for the analysis of the highly inclined events, Auger data is found to lie completely outside the phase space predicted by the models, as can be seen in Figure \ref{fig:comp} (right). So, even for cosmic rays with a primary energy much closer to the maximum center of mass energy achieved in the latest LHC run, a deficit in the production of muons in the hadronic models is found. The complete analysis of UMD data can be found in \cite{UMD_2020}.      

\section{Fluctuations in the number of muons}
\label{sec:fluctus}

\begin{center}
  \begin{figure}
    \begin{subfigure}{0.5\textwidth}
      \includegraphics[width=0.9\textwidth]{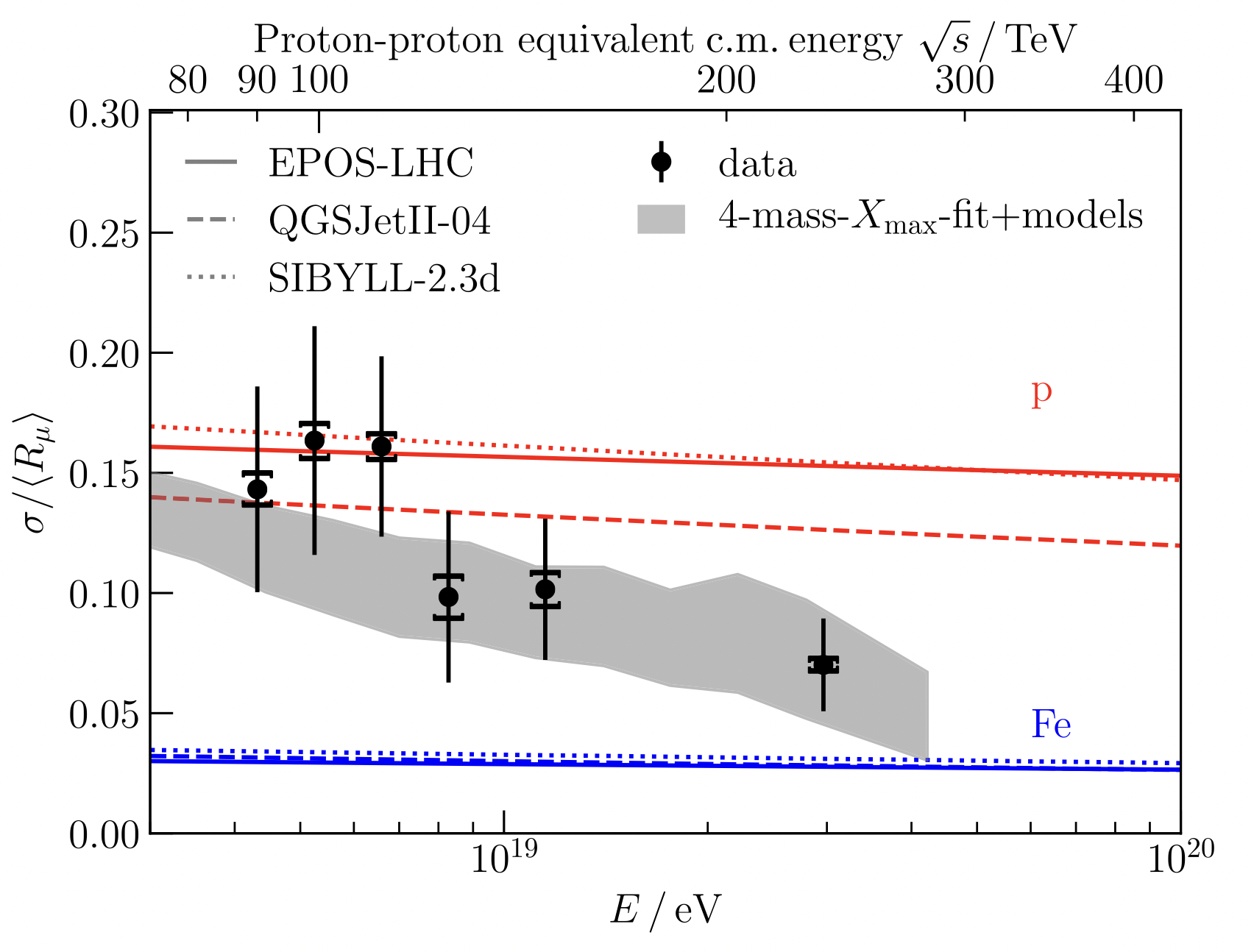}
    \end{subfigure}
    \hfill
    \begin{subfigure}{0.5\textwidth}
      \includegraphics[width=0.9\textwidth]{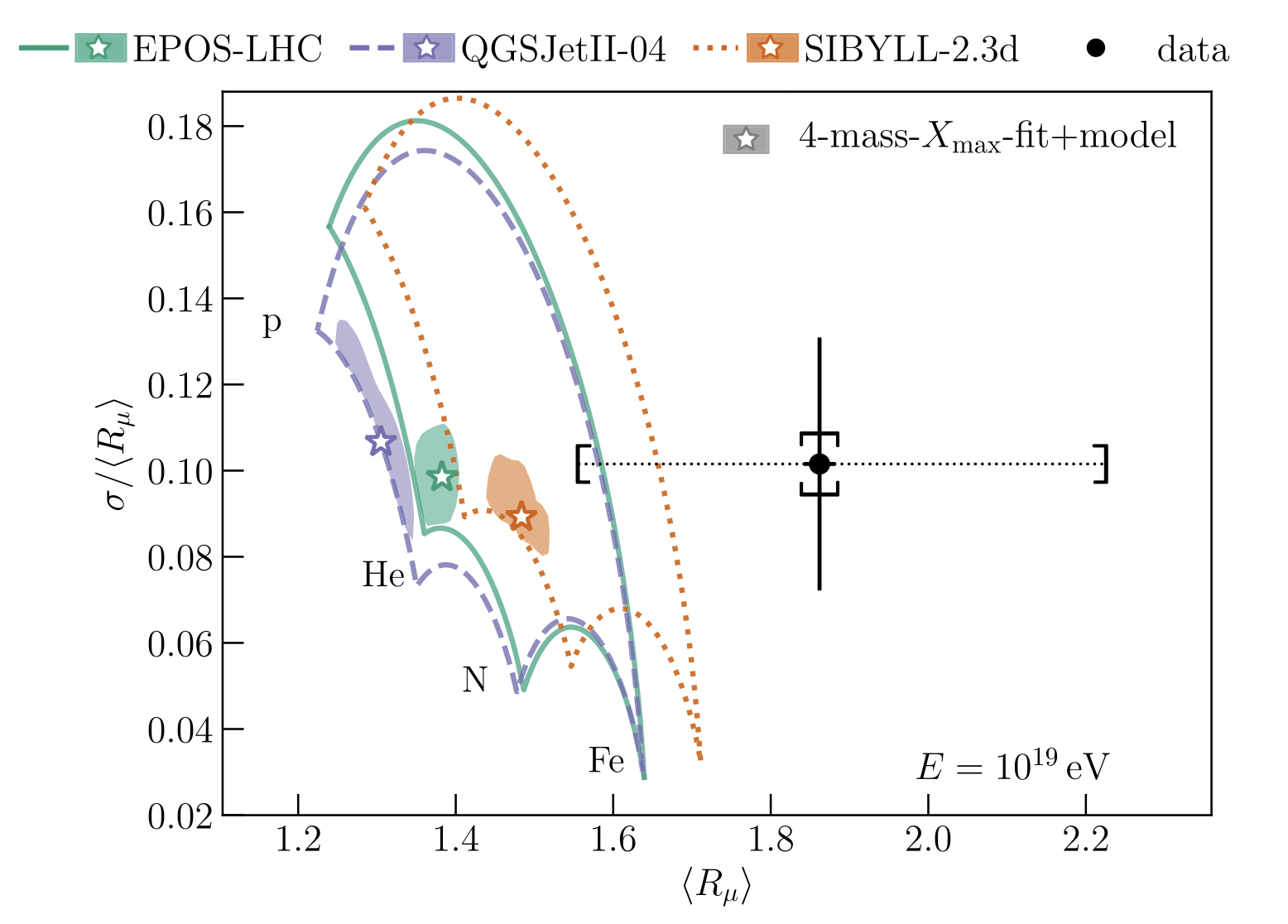}
    \end{subfigure}
    \caption{\emph{Left:} Relative fluctuations in the number of muons as a function of energy. Predictions from hadronic interaction models for proton and iron primaries are also shown. The grey band shows the expected values from mass composition measurements. \emph{Right:} Auger measurement and model predictions for the fluctuations and average muon number for the events in the energy bin corresponding to $10^{19}$ eV\@. \cite{Fluctuations_2021}}
    \label{fig:fluctus}
  \end{figure}
\end{center}

The latest Auger contribution to the muon deficit issue is an analysis of high energy, highly inclined events, in which the fluctuations in the number of muons were considered for the first time. The fluctuations relative to the average muon number are shown as a function of energy in Figure \ref{fig:fluctus} (left). Auger data is well contained between the model predictions and, furthermore, it is compatible with predictions contemplating the mass composition information based on independent \Xmax\, measurements, shown in the plot as a grey band. Figure \ref{fig:fluctus} (right) shows the measured relative fluctuations and the muon number (black marker) for the events in the energy bin corresponding to $10^{19}$ eV\@. The colored contours determine the regions containing the expected values for any combination of the four indicated primaries (p, He, N and Fe) while the star symbols indicate the expected values for the mixture of primaries that results from \Xmax\, measurements and the shaded areas represent their uncertainties. This plot shows simultaneously how the fluctuations in the number of muons are adequately predicted while the average muon number falls short for all models. This result favours a small muon deficit present at every stage of the shower over a large discrepancy only in the first interactions. The details of this analysis and further discussion of its results can be found in \cite{Fluctuations_2021}.  

\section{Conclusion}
\label{sec:conclus}

The hybrid detection techniques that the Pierre Auger Observatory uses to study ultra-high-energy cosmic rays provide insights into the hadronic interactions taking place in the development of extensive air showers. Different analysis have shown that the most updated hadronic interaction models fail to predict the muon content at the ground that is seen with Auger data, when mass composition information is considered. The fluctuations in the number of muons have also been studied and shown to be  adequately described by the models. It all seems to indicate that the muon production mechanism of the models has a small deficit at every stage of shower development, and several theoretical models are currently under discussion to explain and, eventually, fix this issue. The upgrade of the Pierre Auger Observatory will provide further insight into both the primary mass composition and hadronic interactions of ultra-high-energy cosmic rays.  

\bibliography{SciPost_Example_BiBTeX_File.bib}

\nolinenumbers

\end{document}